# Modéliser le physique pour comprendre le contrôle : le cas de l'anticipation en production de parole.

Pascal Perrier

Institut de la Communication Parlée, UMR CNRS 5009,
INPG & Université Stendhal, Grenoble

Avec la collaboration de
Yohan Payan[1] et Romain Marret[2]

[1]: Laboratoire Techniques de l'Imagerie, de la Modélisation et de la Cognition
UMR CNRS 5525, Université Joseph Fourier, Grenoble

[2]: Institut de la Communication Parlée

# 1  Introduction

Pour étudier, comprendre et modéliser le contrôle des gestes de la parole, comme d'ailleurs pour étudier le contrôle moteur des gestes humains en général, le substrat de base est constitué d'une quantité très importante de données physiologiques, cinématiques, acoustiques et éventuellement dynamiques, collectées sur des locuteurs humains. Cette étape de collecte et d'analyse de données a constamment progressé au cours des vingt dernières années, et ceci grâce au développement et à l'utilisation de techniques de mesure efficaces et sophistiquées, telles que l'endoscopie, la transillumination glottique, l'électropalatographie, la électromagnétométrie bi- ou tri-dimensionnelle, l'Imagerie par Résonance Magnétique ou encore les micro-capteurs de pression mécanique (Hardcastle, 1984 ; Perkell *et al.*, 1992 ; Hoole, 1999 ; Masaki, *et al.*, 1999 ; Ziert *et al*, 1999). L'intérêt d'une telle démarche d'investigation expérimentale est indéniable puisqu'elle a permis de déterminer les corrélats physiques, articulatoires et acoustiques essentiellement, des sons de la parole et de leurs enchaînements. Ces travaux ont ainsi fait avancer de manière significative les débats sur les relations entre phonologie et phonétique (voir en particulier : Perkell & Klatt, 1986 ; Stevens: 1972, 1989), ainsi que la typologie des langues (Ladefoged & Maddieson, 1976) ou encore, plus récemment, la synthèse de visages parlants (Elisei *et al.,* 2001 ; Badin *et al.*, 2002, Engwall, 2003).

Cependant, pour comprendre comment la production de la parole est contrôlée, cette démarche, pour indispensable qu'elle soit, ne suffit pas. En effet, le système périphérique de production de la parole a des caractéristiques qui lui sont propres et qui influent sur la façon dont les articulateurs de la parole (mandibule, langue, lèvres, velum, cordes vocales) se déplacent au cours du temps, ainsi que sur la façon dont ils interagissent les uns avec les autres. Nos systèmes de mesure physiologique, cinématique, acoustique ou dynamique, aussi sophistiqués soient-ils, ne permettent donc pas d'observer les conséquences directes des commandes envoyées par le système nerveux central, mais seulement celles de l'interaction entre ces commandes et le système physique de production de la parole. Les incidences de ce constat sur l'interprétation des données recueillies sur des locuteurs sont potentiellement d'importance, ainsi qu'en attestent les 3 exemples suivants.

Considérons en premier lieu les travaux de Nelson (1983) qui a proposé que les caractéristiques de vitesse des mouvements des articulateurs de la parole (mandibule et langue) soient le résultat d'un contrôle optimal dont l'objectif serait de minimiser "l'effort" fourni par le locuteur. Observant les *profils de vitesse* (c'est-à-dire la variation temporelle de la vitesse entre deux passages par zéro de la vitesse) de ces mouvements, il a constaté qu'ils avaient une forme dite "*en cloche*", c'est-à-dire avec un seul maximum et une augmentation puis diminution progressive de part et d'autre de ce maximum. Par simulations, Nelson montra qu'un tel profil de vitesse est typiquement celui qui est obtenu quand le *jerk* (c'est-à-dire la variation de l'accélération au cours du mouvement) est minimal, ce qui l'amena à émettre sa proposition sur un contrôle optimal. Mais il est par ailleurs bien connu qu'un système dynamique linéaire du second-ordre non amorti, du type masse-ressort, a une évolution sinusoïdale de sa vitesse, ce qui donnera aussi, entre deux passages successifs par zéro de la vitesse, un profil de vitesse en cloche, typiquement un arc de sinusoïde. Or, de nombreux travaux (voir par exemple Ostry & Munhall, 1985) ont montré expérimentalement que les mouvements des articulateurs de la parole avaient des caractéristiques de systèmes dynamiques du second ordre et pas seulement lors des tâches de parole. Ces différentes études sont donc, du point de vue des données, cohérents avec les observations de Nelson (1983). Mais les explications sous-jacentes sont différentes : au lieu d'être l'effet d'une stratégie spécifique de contrôle, la forme des profils de vitesse proviendrait en effet des caractéristiques dynamiques intrinsèques des articulateurs. On pourrait aussi envisager que cette caractéristique cinématique soit le résultat de la conjonction des deux phénomènes, les propriétés dynamiques et le contrôle. Au regard des données seules, toutes ces explications apparaissent comme plausibles. La seule manière de trancher est de caractériser les propriétés physiques de la langue ou de la mandibule et d'en inférer leurs influences potentielles sur les profils de vitesse.

Un second exemple, toujours relatif aux profils de vitesse, concerne l'interprétation qui est faite du nombre des maxima de vitesse (classiquement appelés *pics de vitesse*) existant dans un mouvement de parole entre deux sons élémentaires. Les travaux initiés par Adams *et al.* (1993) ont en effet mis en évidence qu'il n'existait en général qu'un ou deux pics de vitesse lorsque le débit d'élocution était rapide, alors qu'il en existait plusieurs à débit d'élocution lent. Les auteurs ont alors suggéré que le passage d'un mouvement lent à un mouvement rapide passe par une réorganisation importante des stratégies de contrôle moteur, les mouvements lents étant ainsi, selon cette hypothèse, produits non plus comme des mouvements d'une cible vers une autre, mais comme des successions de petits mouvements intermédiaires. En utilisant un modèle mécanique bi-dimensionnel de la langue (Payan & Perrier, 1997), qui prend en compte avec un certain degré de réalisme les propriétés physiques de cet articulateur, nous avons pu montrer que le nombre des pics de vitesse n'était pas systématiquement l'image directe du contrôle moteur sous-jacent. En effet, pour des patrons temporels de commandes motrices totalement similaires, les transitions entre voyelles générées par le modèle pouvaient, selon les voyelles, comporter un ou plusieurs pics de vitesse, selon les muscles impliqués et selon l'évolution temporelle des forces à l'origine du mouvement.

Une dernière illustration de la difficulté d'interprétation des données cinématiques est fournie par l'observation, répétée à plusieurs reprises et dans différentes langues, de trajectoires courbes en forme de boucles pour des points placés sur le contour supérieur de la langue au cours de la production de séquences Voyelle-Consonne-Voyelle (VCV) où la consonne était vélaire ([k] ou [g]). C'est Houde (1967) qui, le premier, a décrit ce type de patron articulatoire à partir de l'analyse de films cinéradiographiques des mouvements linguaux pendant des séquences Voyelle1-[g]-Voyelle2 (V1[g]V2). Il notait en particulier que, pendant sa phase de contact avec le palais correspondant à l'occlusion consonantique, la langue se déplaçait vers l'avant, et ceci même si le geste vocalique [V1-V2] était orienté vers l'arrière comme, par exemple, dans la séquence [iga]. La question de l'origine de la forme de ces trajectoires a été l'objet de nombreux débats dans la littérature. Récemment, Löfqvist & Gracco (2002) ont proposé que ces boucles soient la conséquence du fait que les trajectoires sont courbes, et non rectilignes, et, s'inspirant d'études menées sur le contrôle du bras dans des tâches de pointage de cibles (voir par exemple Hogan, 1984), ils ont attribué l'origine de cette courbure à des principes généraux de contrôle moteur qui viseraient à minimiser un certain effort au cours du mouvement. Ainsi donc, selon Löfqvist & Gracco (2002) ces trajectoires en forme de boucle seraient la conséquence indirecte d'une stratégie optimale de contrôle moteur. Cette proposition est plausible, mais elle ne peut être défendue de manière convaincante sur la seule base des signaux de mouvement. En effet, en effectuant des simulations avec un modèle biomécanique de la langue (Payan & Perrier, 1997), nous avons montré que les boucles observées pourraient tout aussi bien être une conséquence directe de la biomécanique de la langue, et éventuellement, dans certains cas, de son interaction avec les forces de pression de l'air dans la cavité arrière du conduit vocal (Perrier et al., 2000). Nous avons en effet trouvé que, quelles que soient V1 et V2, les trajectoires sont courbes, et jamais rectilignes. De plus, lorsque la première voyelle V1 était une voyelle d'arrière ([u] ou [a]), les larges boucles observées expérimentalement s'expliquent parfaitement par la variation temporelle des directions de forces musculaires appliquées à la langue et par l'interaction entre le palais et la langue pendant la phase d'occlusion consonantique (Perrier *et al.*, 2003). Dans le cas où la voyelle V1 est une voyelle d'avant ([i]), si l'interaction entre le flux d'air dans le conduit vocal et la langue est prise en compte grâce au couplage entre le modèle biomécanique et un modèle de flux, on observe des petites boucles vers l'avant, chose que la biomécanique seule ne prédisait pas (Perrier *et al.* 2000).

En conclusion, il existe de nombreux exemples dans la littérature qui confirment le point de vue selon lequel l'interprétation de données expérimentales sur la production de la parole ne peut pas se faire de manière convaincante sans prendre en compte les phénomènes physiques sous-jacents.

Dans cet article, nous allons nous intéresser au phénomène d'anticipation en production de la parole, tel qu'il a pu être observé par le passé sur les mouvements des articulateurs du conduit vocal, et, en exploitant le modèle biomécanique de la langue (Payan & Perrier, 1997) déjà mentionné ci-dessus, nous évaluerons dans quelle mesure il pourrait être imputable à des phénomènes purement biomécaniques.

## 2  Quelques notions sur l'anticipation en production de parole.

L'étude de la production de la parole s'appuie sur le postulat que la production d'une chaîne de phonèmes est fondamentalement une tâche séquentielle consistant à produire le phonème (i) avant le phonème (i+1) et après le phonème (i-1). On parle alors d'anticipation lorsqu'on trouve soit dans les signaux articulatoires, soit dans le signal acoustique, des indices sur les caractéristiques articulatoires du phonème (i+1) avant qu'on ait atteint la réalisation du phonème (i).

De nombreux travaux expérimentaux ont mis en évidence des phénomènes d'anticipation et ont contribué ainsi aux grands débats sur le contrôle de la production de la parole. Les plus fameux d'entre eux, parce qu'ils ont débouché sur des propositions importantes concernant la modélisation du contrôle de la production de la parole, sont ceux de Henke (1966) (modèle *look-ahead*, utilisé pour le français par Benguerel & Cowan, 1974), Bell-Berti et collègues (Bell-Berti & Harris; 1981 ; Boyce *et al*, 1990) (modèle *time-locked*), Perkell (1990) (modèle *hybride*), et Abry & Lallouache (1991 ; 1995) (*Modèle de l'Expansion du Mouvement*). Dans tous les cas, les observations expérimentales ont clairement démontré que le geste articulatoire sous-jacent à la production d'un son élémentaire à l'intérieur d'une chaîne phonétique donnée, pouvait commencer bien avant que le son élémentaire précédent n'ait été émis. Mais les contrôles sous-jacents proposés pour expliquer ces données diffèrent selon les auteurs. Sans rentrer dans les détails de ces propositions, qui ne seront pas débattues dans cet article, rappelons pour résumer quelles en sont les grandes tendances. Pour le modèle *look-ahead*, les traits articulatoires caractérisant un phonème se rétro-propagent aux phonèmes précédents tant qu'ils sont compatibles avec les traits caractéristiques de ces derniers ; c'est donc un modèle que l'ont pourrait qualifier de "*tout phonologique*", dans la lignée de l'école linguistique du MIT, où les caractéristiques propres du système de production sont résolument ignorées. A l'opposé; le modèle *time-locked,* dans la tradition "*tout dynamique*" des Laboratoires Haskins, met l'accent sur les contraintes dynamiques des articulateurs de la parole, et propose que la durée de l'anticipation soit déterminée par le temps de réponse des articulateurs à une commande gestuelle, et ceci afin que la configuration articulatoire requise soit atteinte au moment où le son associé est émis. Le modèle *hybride* est, comme son nom l'indique, un mélange des deux modèles précédents ; ainsi selon lui le mouvement commencerait bien aussi tôt que la description phonologique l'autorise, mais la dernière partie du mouvement vers la position cible serait essentiellement déterminée par les caractéristiques dynamiques des articulateurs mis en jeu. Enfin le *Modèle d'Expansion du Mouvement* propose que, pour un geste articulatoire donné, la durée de l'anticipation augmente linéairement avec le temps séparant deux sons successifs pour lesquels ce geste articulatoire est pertinent, et ceci selon une stratégie propre à chaque locuteur ; la durée d'anticipation peut ainsi, pour certains locuteurs, atteindre celle que prédirait le modèle *look-ahead,* mais ce n'est pas non plus systématiquement le cas ; cette durée d'anticipation ne peut par ailleurs pas être inférieure à un seuil minimum, en l'occurrence la durée d'anticipation qui est mesurée quand deux sons, pour lesquels le geste articulatoire considéré est pertinent, se succèdent immédiatement.

Le rôle que jouent ces phénomènes d'anticipation dans l'interaction locuteur/auditeur a par ailleurs été clairement mis en évidence par un ensemble de tests perceptifs, et ceci à plusieurs occasions, aussi bien dans le domaine auditif (Benguerel & Adelman, 1976 ; Hirsch *et al.*, 2003 ; Roy et al., 2003 ; Vaxelaire *et al.*, 2003) que dans le domaine visuel (Cathiard, 1994 ; Cathiard *et al.*, 1998). Ceci suggère que ces phénomènes d'anticipation constituent une partie importante de la caractérisation physique de la parole : le locuteur les produit, et l'auditeur sait les interpréter pour savoir ce qui va venir ; on peut donc penser qu'ils constituent même une partie de la représentation que locuteur et auditeur ont d'une séquence phonétique donnée. En dépit des propositions d'adaptation proposées par Boyce *et al.* (1990), nous ne pensons pas que le modèle *time locked,* qui ne prend pas en compte la possibilité d'un contrôle central d'au moins une partie de ces phénomènes d'anticipation, puissent rendre compte des durées d'anticipation observées par exemple par Benguerel & Cowan (1974) ou Abry & Lallouache (1995). Nous restons convaincus que de telles durées ne peuvent être obtenues que si une large partie des phénomènes d'anticipation est le résultat d'une planification centrale. Cependant, rejoignant en cela Bell-Berti & Harris (1981), nous n'écartons pas *a priori* la possibilité que les caractéristiques physiques du système périphérique de production de la parole puissent contribuer à une partie de l'anticipation mesurée sur les signaux articulatoires et acoustiques de la parole.

Cette hypothèse est confirmée par les travaux de Ostry et collègues (Ostry *et al.*, 1996 ; Perrier *et al.*, 1996a) qui ont évalué les contributions potentielles des caractéristiques mécaniques de la mandibule à l'anticipation mesurée sur les signaux de mouvements mandibulaires dans des séquences du type Voyelle-Consonne plosive – Voyelle (VCV dans la suite) impliquant un mouvement d'élévation puis d'abaissement de la mandibule. Ces auteurs ont en effet comparé des données cinématiques sur les mouvements de la mandibule mesurées sur des locuteurs au cours de la production de la parole et des simulations obtenues avec un modèle biomécanique de la mandibule (Laboissière *et al.*, 1996) dont les mouvements étaient générés, selon la *Théorie du Point d'Equilibre* de Feldman (1986), par le déplacement contrôlé de son état d'équilibre entre des positions cibles associées aux différents phonèmes de la séquence VCV considérée. Ils ont ainsi montré, en faisant de nombreuses simulations dont les commandes motrices ne différaient l'une de l'autre que dans la séquence CV, les commandes relatives à la séquence VC restant strictement identiques, que des effets d'anticipation existaient qui modifiaient la trajectoire du modèle dans la séquence VC (celle dont les commandes motrices n'étaient pas modifiées), et ceci de manière qualitativement comparable aux données mesurées sur des locuteurs. Ils montraient ainsi, pour la première fois de manière explicite, que, sans qu'aucune anticipation n'existe au niveau de la commande centrale, des effets d'anticipation pouvaient être observés sur les données cinématiques. Dans le cadre du modèle proposé, ces phénomènes, que nous qualifierons de *d'anticipation périphérique* parce que mesurés sur les signaux cinématiques mais pas sur les signaux de contrôle, étaient essentiellement dus à l'inertie du système et à la façon dont la force musculaire était générée.

Ces travaux nous ont incités à poursuivre dans cette voie, mais cette fois-ci non plus sur la mandibule mais sur la langue, articulateur hautement déformable dont les caractéristiques inertielles et mécaniques, et la fonction dans la production de la parole, diffèrent grandement de celles de la mandibule.

# 3 Anticipation en parole et caractéristiques physiques de la langue.

## 3.1 Modélisation biomécanique de la langue

Pour cette étude, nous avons utilisé un modèle biomécanique bi-dimensionnel de la langue dont les caractéristiques ont été décrites en détail dans Payan & Perrier (1997) et Perrier *et al*, (2003). Nous n'en rappellerons donc ici que les grandes lignes, et incitons le lecteur à se reporter à ces deux publications pour plus de détails.

Il s'agit d'une modélisation à éléments finis, dont les caractéristiques élastiques ont été adaptées à partir de données sur les muscles rapides (par exemple le muscle cardiaque) de façon à ce que des mouvements d'amplitudes réalistes puissent être générés avec des vitesses tangentielles réalistes et avec des forces musculaires de l'ordre de quelques newtons. La forme au repos de la structure à éléments finis a été adaptée de façon à représenter fidèlement la forme de la langue d'un locuteur de langue française en position neutre correspondant globalement à un schwa. Chaque muscle est représenté dans la structure à éléments finis par un sous-ensemble d'éléments spécifique dont les caractéristiques élastiques sont susceptibles de varier lorsque le muscle est recruté. Les forces musculaires sont appliquées en certains nœuds de la structure à éléments finis *via* des macro-fibres dont les directions principales sont conformes aux données anatomiques sur la structure musculaire de la langue. Tout comme dans le modèle de mandibule utilisé par Ostry et collègues (cf. section 2.), le mécanisme de génération et de contrôle de la force musculaire est inspiré de la *théorie du point d'équilibre* de Feldman (1986). Six muscles sont modélisés : le génioglosse que nous divisons en deux entités indépendantes, postérieure et antérieure, le hyoglosse, le styloglosse, le longitudinal supérieur, le longitudinal inférieur, et le vertical. Ils permettent de rendre compte correctement des principales déformations de la langue dans le plan sagittal.

Pour une séquence de phonèmes donnée, des positions d'équilibre cibles sont définies pour chaque phonème, et le mouvement est produit entre ces positions cibles par la variation à vitesse constante des variables de contrôle spécifiant cet état d'équilibre. Les variables temporelles des commandes motrices spécifient la durée de la transition entre positions cibles ainsi que la durée de tenue de ces positions.

Précisons enfin que dans la version du modèle que nous avons utilisée pour ce travail, la mandibule est fixe.

## 3.2 Evaluation de l'influence de caractéristiques biomécaniques de la langue sur l'anticipation dans les mouvements linguaux en parole.

Pour évaluer l'influence potentielle des caractéristiques biomécaniques de la langue sur l'anticipation mesurée sur les mouvements de la langue en parole, nous avons généré des séquences Voyelle1-Voyelle2-Voyelle 3 (V1-V2-V3 dans la suite) et des séquences V1-C-V2. Pour ne mesurer que les effets potentiels de la biomécanique de la langue la position cible adoptée pour un phonème donné a été la même pour toutes les séquences, et ceci quels que soient les autres phonèmes de la séquence considérée. En d'autres termes, dans cette phase du travail nous n'avons pris en compte aucun effet potentiel de planification centrale.

### 3.2.1 Séquences V1-V2-V3

Nous avons fait un ensemble de simulations où les commandes motrices de la séquenceV1-V2 ont été maintenues constantes pendant que celles V3 variaient d'une simulation à l'autre.

Nous avons alors observé si la trajectoire de la langue pour la séquence V1-V2 était affectée par la variation de V3. Trois conditions temporelles ont été testées : dans la condition 1, le temps de transition entre positions cibles était égal à 50 ms et le temps de tenue des cibles à 150 ms ; dans la condition 2, le temps de transition entre positions cibles était égal à 30 ms et le temps de tenue des cibles à 80 ms, afin de simuler un faible ratage de cibles (cf. Loevenbruck & Perrier, 1993 ; Perrier *et al.*, 1996b) ; dans la condition 3, le temps de tenue des cibles est encore plus réduit (40 ms) de façon à provoquer un ratage de cible plus net.

Les résultats que nous avons ainsi obtenus sont illustrés par les figures 2 à 4. Elles représentent les déplacements horizontaux et verticaux d'un point situé sur le contour supérieur du modèle de langue, dans la région palatale (cf. Figure 1), ainsi que sa vitesse tangentielle, et ceci pour les séquences [schwa-a-e-a] *versus* [schwa-a-e-i], dans les 3 conditions temporelles évoquées ci-dessus.. Même si, évidemment, les caractéristiques cinématiques ne sont pas les mêmes en tous les points de la langue, ce point est bien représentatif des effets globaux que nous avons observés. Dans ces trois figures, la langue est au départ dans sa position de repos (schwa). La langue recule ensuite (la variable horizontale x croît) et s'abaisse (la variable verticale y décroît) vers le [a]. Selon la condition temporelle, la position du [a] est maintenue plus ou moins longtemps, puis la langue avance (x décroît) et remonte (y croît) vers le [e] qui à son tour est maintenu plus ou moins longtemps selon la condition temporelle. Enfin, selon la nature de la troisième voyelle, la langue va reculer et s'abaisser (vers le [a]) ou continuer son mouvement d'avancée et d'élévation (vers le [i]). Les différences entre ces trois figures seront commentées dans la section 3.2.3 ci-dessous.

### 3.2.2 *Séquences V1-[k]-V2*

Nous avons ensuite fait, pour les trois mêmes conditions temporelles, un ensemble de simulations incluant la consonne plosive [k]. Il convient de préciser ici que dans le modèle de contrôle que nous avons adopté, les consonnes sont définies comme les voyelles, c'est-à-dire par une position d'équilibre cible. Mais, à la différence des voyelles qui peuvent atteindre cette position d'équilibre cible si les contraintes dynamiques et temporelles le permettent, la consonne plosive ne peut jamais l'atteindre car la langue est stoppée en chemin par le palais. Les positions d'équilibre cibles associées aux consonnes plosives sont donc des cibles virtuelles et inaccessibles. C'est ainsi que, dans notre modèle de contrôle, le contact avec le palais est réalisé, sur la surface voulue et pour la durée voulue (Fuchs *et al.*, 2001). Rappelons aussi que si les ratages de cibles sont possibles avec les voyelles sans que la perception du signal acoustique soit nécessairement mise en danger, ils ne peuvent pas exister avec des consonnes plosives, pour lesquelles le contact avec le palais est indispensable et ceci pour une durée suffisante.

Nous présentons ici les résultats des simulations pour les séquences [schwa-a-k-e] *versus* [schwa-a-k-ɔ] (cf. Figures 5 à 7). Comme dans les séquences de voyelles, la langue est au départ dans sa position de repos, puis elle recule et s'abaisse pour atteindre le [a].. Elle monte ensuite pour aller produire le [k] qui est marqué par un long plateau selon la direction y dans les conditions temporelles 1 et 2. Ce plateau, plus long que celui que nous observions pour les voyelles sur les figures 2 et 3, est dû au contact entre la langue et le palais qui bloque l'ascension de la langue vers sa cible virtuelle. Puis selon la voyelle suivante, la langue va s'abaisser en allant soit vers l'avant ([e]) soit vers l'arrière ([ɔ]).Les différences entre ces trois figures seront commentées dans la section 3.2.3 ci-dessous.

### 3.2.3 *Analyse des résultats*

Dans la condition temporelle 1, on observe pour la séquence V1-V2-V3 des plateaux sur les signaux de position tant dans la direction horizontale que dans la direction verticale. Les positions cibles des différentes voyelles sont donc bien atteintes. On n'observe aucun effet d'anticipation : les positions selon x et y, et la vitesse tangentielle sont en effet rigoureusement identiques jusqu'à l'instant où la langue quitte la position cible du [e] (au temps 0.415). Dans les conditions temporelles 2 et 3, on n'observe plus de plateaux sur les signaux de position, et les positions atteintes pour les voyelles [a] et [e] sont plus proches de la position de repos. On assiste à un phénomène de ratage de cible, et ceci de manière d'autant plus marqué que la durée de la tenue des cibles est réduite (de la condition 2 à la condition 3). Ces phénomènes de ratage de cible s'accompagnent d'une séparation claire des signaux de position selon y avant que la position la plus haute du [e] n'ait été atteinte. Dans la condition 2, la position la plus haute du [e] est en effet atteinte à l'instant 0.210 alors que les signaux se séparent dès l'instant 0.195. Ceci est encore plus marqué dans la condition 3 avec respectivement les instants 0.155 pour le [e] et 0.135 pour la séparation. Dans les deux conditions, le signal de position de la séquence [schwa-a-e-i] est plus haut que celui de la séquence [schwa-a-e-a]. Cette différentiation qui intervient dans la transition [ae] pourrait donc être interprétée comme le signe d'une anticipation de la production de la voyelle V3. En l'occurrence il n'en est rien, puisque les commandes motrices cibles du [a] et du [e] sont rigoureusement les mêmes pour les deux séquences. Nous assistons donc à un phénomène "d'anticipation périphérique" observable sur les signaux cinématiques, mais non planifié au niveau des commandes motrices.

Pour ce qui concerne les simulations du type [V1-C-V2], on constate qu'aucune séparation nette des signaux de position n'existe avant que la langue ne s'abaisse à la suite du contact consonantique entre la langue et le palais. On observe certes, dans les conditions 2 et 3, de petites différences entre les signaux de position selon la direction x, mais elles sont très faibles et ne constituent en aucun cas une séparation : la différence ne va pas dans la direction des positions caractéristiques selon x des voyelles subséquentes. Selon nous cette absence "d'anticipation périphérique" est associée au fait que, même dans la condition 3, le contact entre la langue et le palais existe. Il n'y a donc pas de ratage de cible, même s'il semble que la durée du contact dans la condition 3 soit trop brève pour correspondre à une production correcte de consonne plosive.

Ainsi les simulations font apparaître que l'influence des caractéristiques biomécaniques de la langue sur la lecture cinématique que l'on peut faire des phénomènes d'anticipation, se limite au cas où des phénomènes de ratage de cible apparaîtraient. Et dans ce cas, ce que nous appelons l'"anticipation périphérique" s'avance au plus jusqu'à la transition vers le son qui précède celui vers lequel cette anticipation est dirigée (dans nos exemples V2). Dans le cas des consonnes, où aucun ratage de cible ne peut être toléré, il n'existe donc pour la langue aucun phénomène d'"anticipation périphérique". Cela peut sembler, à première vue, en contradiction avec les résultats obtenus pour la mandibule par Ostry et collègues (Ostry *et al.*, 1996 ; Perrier *et al.*, 1996a). Mais l'observation attentive des simulations proposées par ces auteurs montrent que les phénomènes d'anticipation qu'ils ont simulés, étaient effectivement associés au ratage de la cible consonantique : c'est parce que la mandibule va moins haut qu'elle commence plus tôt à s'abaisser vers la voyelle subséquente. Ce ratage de cible consonantique, impossible pour la langue dans le cas d'une production consonantique correcte, est par contre tout à fait possible pour la mandibule qui est moins contrainte en position que la langue car, contrairement à cette dernière, elle ne donne pas au conduit vocal sa forme précise. Nos résultats et ceux de Ostry et collègues sont donc parfaitement compatibles. Ils tendent tous deux à montrer le rôle majeur du phénomène de ratage de cible dans l'existence d'un phénomène d'"anticipation périphérique".

Dans tous les cas, les simulations générées avec le modèle de langue, tout comme celles qu'ont publiées Ostry et collègues, tendent à montrer que le phénomène "d'anticipation périphérique" est limité en amplitude et que son empan temporel est faible. Il est donc vraisemblable que les données cinématiques en parole montrant des phénomènes d'anticipation importants en amplitude et en durée sont essentiellement le résultat d'une stratégie de contrôle de haut niveau. Dans la partie 4, nous proposons d'évaluer une version préliminaire d'un modèle de contrôle qui verrait l'anticipation comme une des conséquences naturelles d'une stratégie de contrôle optimal de la production de la parole visant à minimiser l'effort du locuteur tout en assurant la qualité perceptive du son produit.

# 4 Anticipation en parole : le résultat d'une stratégie de contrôle optimal ?

## 4.1 Implémentation du contrôle optimal

La stratégie de contrôle optimal que nous avons implémentée s'inspire de propositions désormais classiques dans le domaine du contrôle moteur, qui reposent sur les deux postulats suivants (Kawato *et al.*, 1990 ; Jordan & Rumelhart, 1992) :

(1) La planification d'une séquence gestuelle consiste à rechercher la séquence de commandes motrices qui produit le mouvement désiré en optimisant un certain critère ;

(2) L'optimisation de ce critère met en jeu des *modèles internes* des relations entre les commandes motrices et les variables physiques de la tâche motrice à exécuter.

### 4.1.1 Elaboration d'un modèle interne statique

La question de la complexité des modèles internes nécessaire à cette planification est un enjeu important de la modélisation du contrôle moteur, et elle est toujours au centre de nombreux débats actuels (voir par exemple la controverse entre Gomi & Kawato, 1996 et Gribble et al., 1998). Dans ce contexte, nous avons adopté le parti pris de la parcimonie, et avons choisi d'exploiter dans notre modélisation de la planification en parole, un modèle interne statique des relations entre les commandes motrices du modèle de langue et les caractéristiques spectrales du signal acoustique de parole produit à partir de la forme de langue ainsi générée et pour des lèvres ouvertes (du type /i/ ou /a/). Nous avons développé par ailleurs (Perrier, 2003), pour le contrôle de la parole, les arguments en faveur de ce modèle interne statique par opposition à des modèles plus complexes tels que ceux que proposent Kawato et collègues (Kawato et al., 1990).

Par *modèle interne statique*, nous entendons un modèle des relations entre valeurs des commandes motrices pour une position cible de la langue et fréquences des maxima du spectre du signal de parole associé (les *formants*). C'est donc un modèle qui n'intègre aucune information cinématique (maximum de vitesse, ou accélération) ni dynamique (mécanismes de génération des forces). Pour constituer ce modèle interne, nous avons procédé en plusieurs étapes (Marret, 2002). La première a consisté à générer l'aire du conduit vocal (on parle de *fonction d'aire du conduit vocal*) à partir de la forme bidimensionnelle du modèle de langue dans le plan sagittal de la tête. Pour cela nous avons utilisé le modèle de Perrier *et al.* (1992) qui repose sur l'analyse d'un moulage du conduit vocal d'un cadavre et sur des mesures par scanner à rayons X effectuées sur un locuteur de langue française qui prononçait les voyelles /i, a, u/. Nous avons arbitrairement considéré le cas d'une aire aux lèvres de l'ordre de $3cm^2$. Ceci permet de générer toutes les voyelles ayant des lèvres ouvertes (pour le français /i/, /e/, / ɛ/, /a/, /ɔ/,/œ/), et nous nous sommes limités à ces cas-ci dans le cadre de la présente étude. Pour chaque forme du modèle de langue dans le plan sagittal de la tête, il a été ainsi possible de générer une forme tri-dimensionnelle du conduit vocal (la *fonction d'aire du conduit vocal*). Dans une seconde étape, cette fonction d'aire a été utilisée comme variable d'entrée d'un modèle harmonique du conduit vocal (Badin & Fant, 1984) qui calcule les formants du signal de parole à partir d'une forme du conduit vocal. Ainsi il a été possible de connaître pour tout jeu de commandes motrices cibles les formants du signal de parole associé, en nous limitant, nous le répétons, au cas d'une aire aux lèvres de 3 cm$^2$. Nous avons alors généré 3000 formes du modèle de langue et les 3000 valeurs des 3 premiers formants (F1,F2,F3) associés. L'étape finale a consisté à apprendre les relations entre les commandes motrices et les valeurs (F1,F2,F3) correspondantes et à les généraliser en utilisant un réseau de neurones à fonctions radiales de base (Poggio & Girosi, 1989). Nous avons ainsi généré un modèle fonctionnel qui, sans nécessiter la résolution des équations du mouvement du modèle biomécanique de la langue, fournit pour un jeu de commandes motrices cibles une bonne approximation des 3 premiers formants du signal de parole correspondant.

### *4.1.2 Génération de séquences optimales*

Le modèle interne statique ainsi constitué a été exploité pour générer des séquences de voyelles (à lèvres ouvertes, nous l'avons dit). Pour cela il faut trouver, pour chaque séquence de voyelles, la séquence des commandes motrices cibles qui minimise un coût que nous avons défini selon les principes suivants. Ce coût doit permettre de prendre en compte deux critères (Marret, 2002). On qualifiera le premier critère de "*orienté vers le locuteur*", car il doit correspondre à une certaine prise en compte de l'effort du locuteur. Pour ce travail, et avant d'envisager des critères plus complexes et probablement plus réalistes, nous avons choisi de minimiser, sur l'ensemble de la séquence, le chemin parcouru par les commandes cibles dans l'espace des commandes motrices. Le second critère est "*orienté vers l'auditeur*". Il a pour objectif de garantir que l'effet perceptif du son produit est correct. Pour cela nous avons défini, sous forme d'ellipses de dispersion dans les plans (F1,F2) et (F2,F3), des zones de l'espace acoustique correspondant à des réalisations perceptivement correctes de chacune des voyelles à lèvres ouvertes du français. Nous avons ensuite défini un critère perceptif qui, pour chaque voyelle, prend une valeur très faible tant que le patron (F1,F2,F3) reste dans la zone de l'espace acoustique de cette voyelle, et qui devient très vite très grand si on s'éloigne de cette zone. Le coût global à minimiser est la somme de ce critère perceptif et de la distance parcourue par les commandes cibles dans l'espace des commandes motrices. Pour atteindre cette minimisation à l'aide du modèle interne statique décrit plus haut, l'algorithme du gradient a été utilisé.

## 4.2 Résultats

La figure 8 présente les résultats de la simulation des séquences [schwa-a-ɛ-i] *versus* [schwa-a-ɛ-ɔ] obtenue à l'aide de ce modèle de contrôle optimal. On note que l'anticipation de la voyelle V3 se traduit par une séparation très marquée des signaux de position aussi bien selon la direction x que selon la direction y. De plus cette séparation intervient non pas à la fin de la transition de [a] vers [ɛ] (c'est-à-dire autour de l'instant 0.480) comme c'était le cas lorsque on observait une "anticipation périphérique", mais dès la fin de la voyelle [a] (au temps 0.23). Comparée à l'influence des caractéristiques biomécaniques de la langue montrée dans la section 3, l'anticipation planifiée selon le modèle de contrôle optimal proposé a donc des incidences beaucoup plus importantes tant du point de vue de la durée de l'anticipation que de l'amplitude de son influence sur les signaux de position. Cela est dû au fait que la planification modifie les commandes motrices cibles correspondant à chaque phonème de la séquence de façon à minimiser le coût décrit ci-dessus.

# 5 Conclusions

Les simulations que nous avons effectuées avec un modèle biomécanique bi-dimensionnel de la langue, ont permis une évaluation de l'influence potentielle des caractéristiques physiques de cet articulateur sur les phénomènes d'anticipation mesurés sur les signaux acoustiques ou articulatoires de la parole. Nos conclusions majeures sont :
  (1) Cette influence est limitée au cas où un phénomène de ratage de cible existerait, c'est-à-dire à la production de séquence de voyelles, puisque la position de la langue est fortement contrainte lors de la production des consonnes.
  (2) Cette influence, quand elle existe, ne se rétropropage pas au-delà de la voyelle qui précède le son vers lequel l'anticipation est dirigée, et son amplitude reste faible.

En conséquence il semble bien que dans l'immense majorité des cas l'anticipation mesurée soit la conséquence de stratégies de contrôle de haut niveau. Nous proposons que le contrôle de cette anticipation n'ait pas de caractère spécifique et que ce phénomène soit le résultat d'une stratégie de contrôle générale visant à minimiser l'effort du locuteur tout en préservant la qualité perceptive du son produit. Les résultats que nous avons obtenus avec une version préliminaire d'un modèle de ce type de contrôle optimal vont dans le sens de notre hypothèse. Il reste à l'évaluer plus précisément en les confrontant plus en détails aux nombreuses données sur l'anticipation en parole. Il sera en particulier important de voir comment un tel modèle sera capable de rendre compte de la variabilité interlocuteur mise en évidence entre autres par Abry & Lallouache (1995).

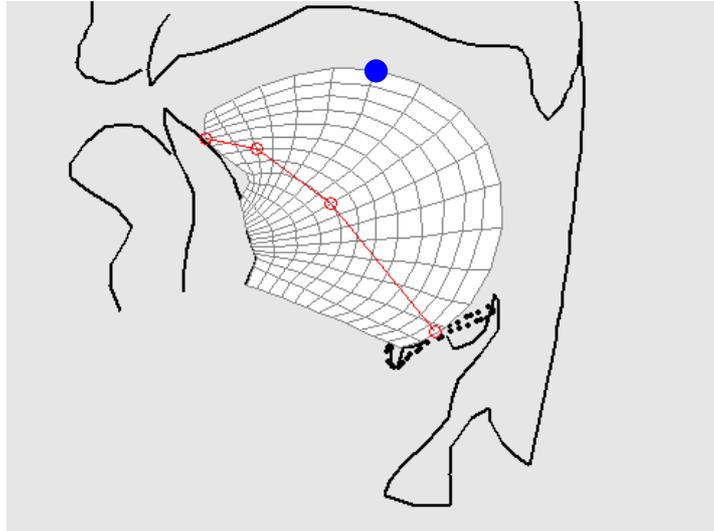

**Figure 1 :** Point sur le contour de la langue (cercle plein) dont les mouvements seront représentés dans les figures suivantes. Cette vue représente la coupe sagittale du conduit vocal du modèle. Les traits noirs représentent les contours pharyngaux (à droite) et palataux (en haut), les lèvres supérieure et inférieure, le contour de l'incisive centrale (en bas à gauche), l'os hyoïde (en pointillé), l'épiglotte et le haut larynx (en bas à droite). Les traits grisés représentent le modèle à éléments finis de la langue.

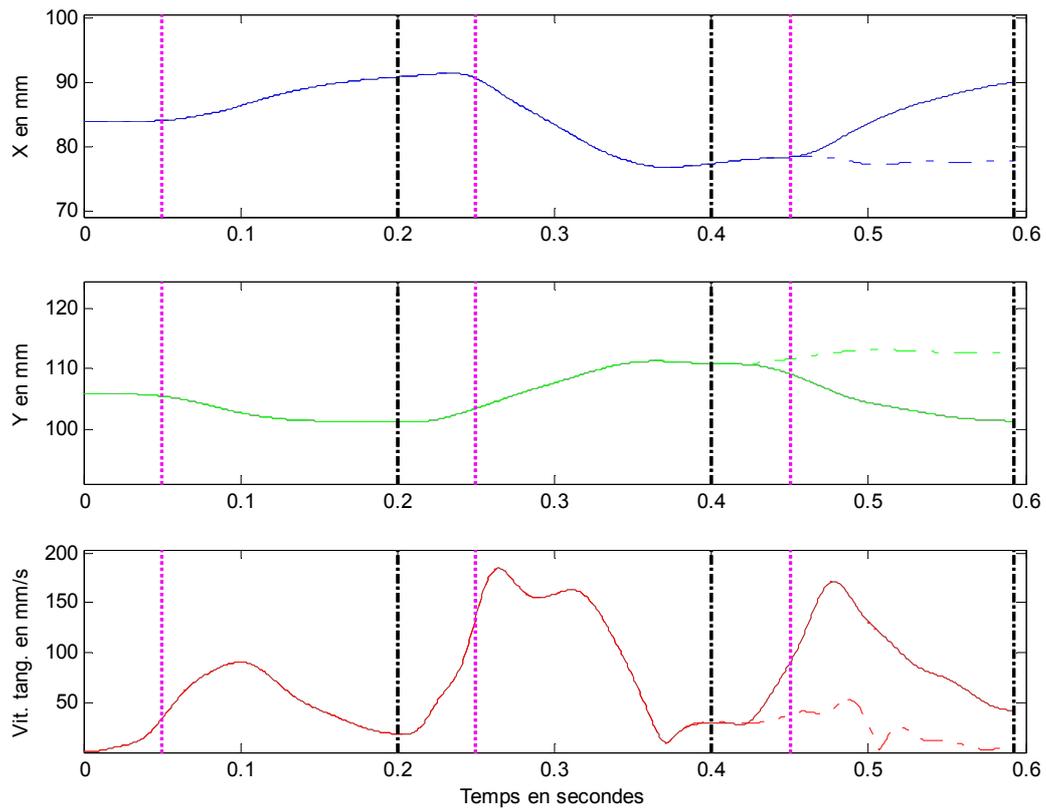

**Figure 2 : Déplacements horizontaux (en haut) et verticaux (au milieu), et vitesse tangentielle (en bas) du point du contour supérieur du modèle représenté à la figure 1 pour les séquences [schwa-a-e-a] (trait continu) et [schwa-a-e-i] (trait tireté) pour la condition temporelle 1. Les traits verticaux pointillés marquent le début de la tenue des commandes motrices cibles de chacun des trois phonèmes autres que le schwa ; les traits verticaux tiretés en marquent la fin.**

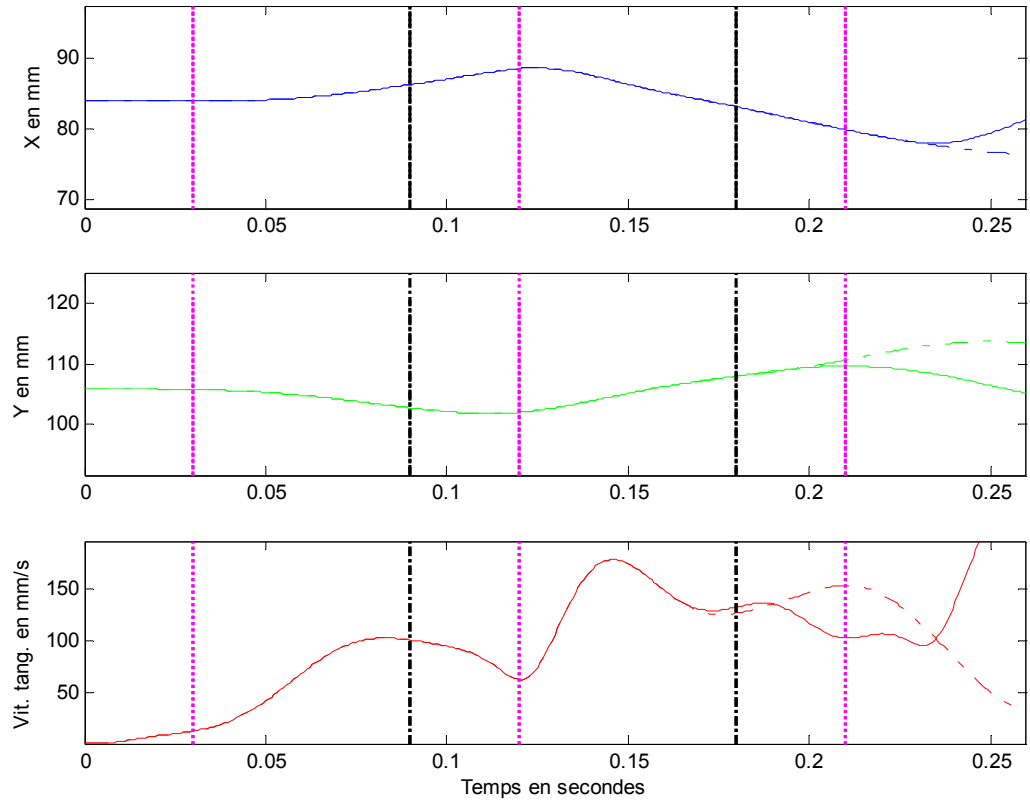

**Figure 3 : Même chose que sur la figure 2, mais pour la condition temporelle 2.**

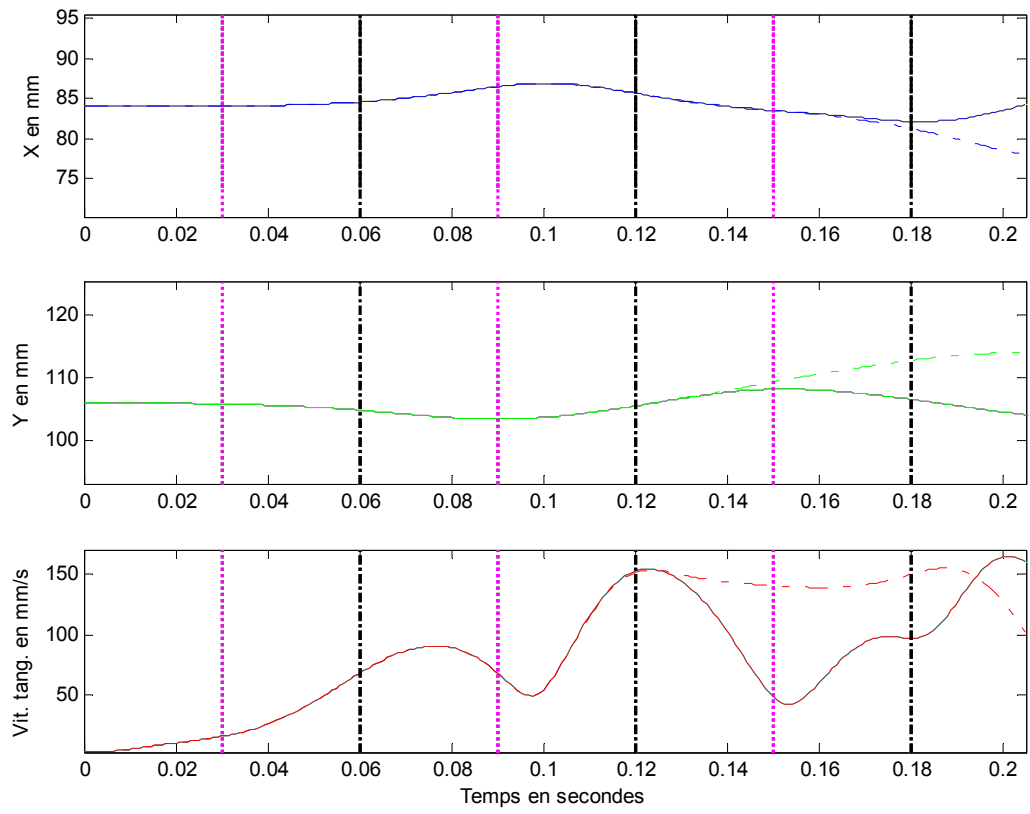

**Figure 4 : Même chose que sur la figure 2, mais pour la condition temporelle 3.**

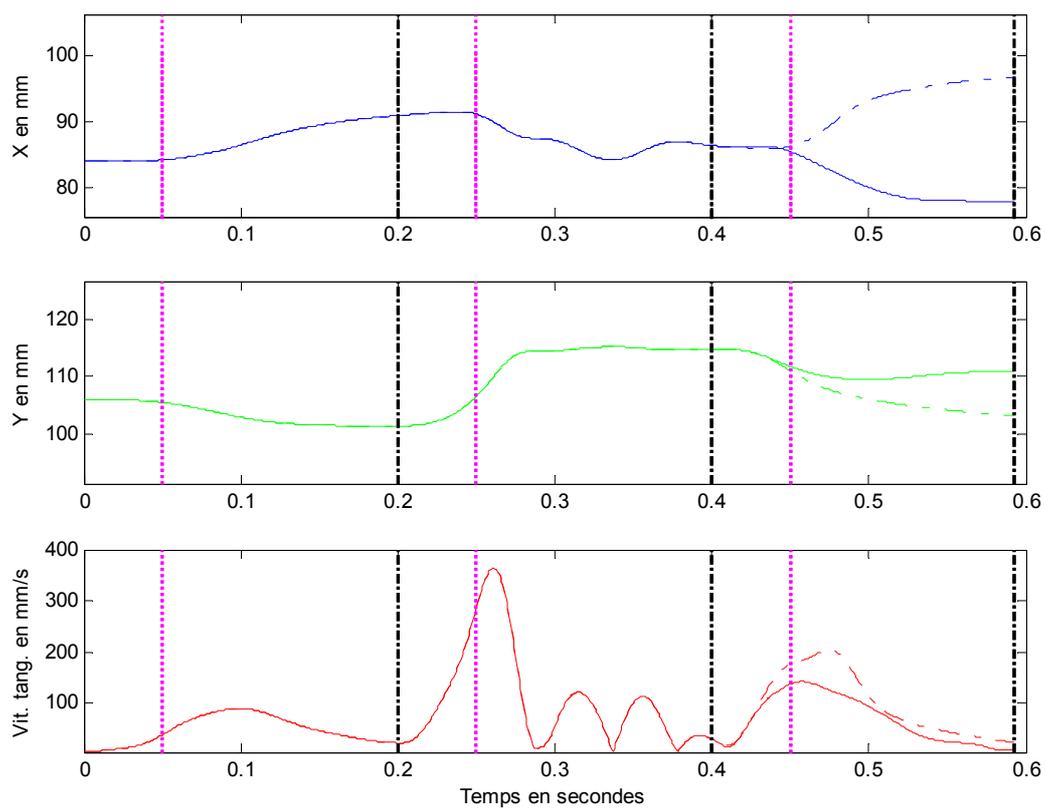

**Figure 5 : Déplacements horizontaux (en haut) et verticaux (au milieu), et vitesse tangentielle (en bas) du point du contour supérieur du modèle représenté à la figure 1 pour les séquences [schwa-a-k-e] (trait continu) et [schwa-a-k-ɔ] (trait tireté) pour la condition temporelle 1.
Pour plus de détails voir figure 2**

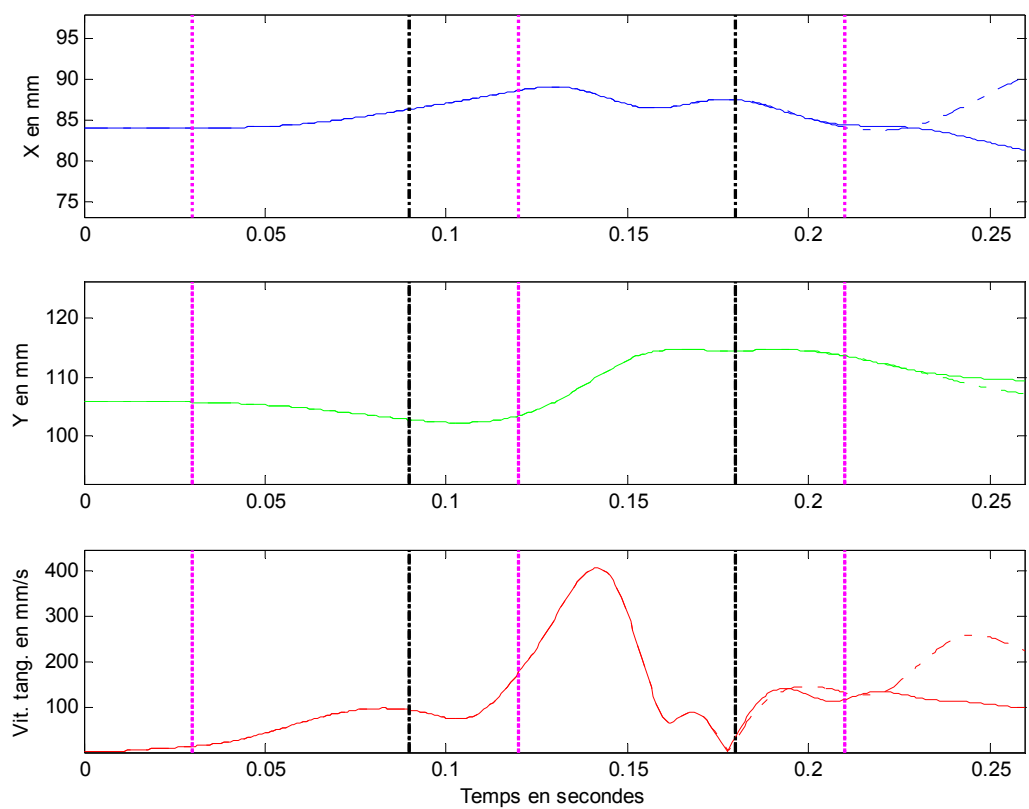

**Figure 6 : Même chose que sur la figure 5, mais pour la condition temporelle 2.**

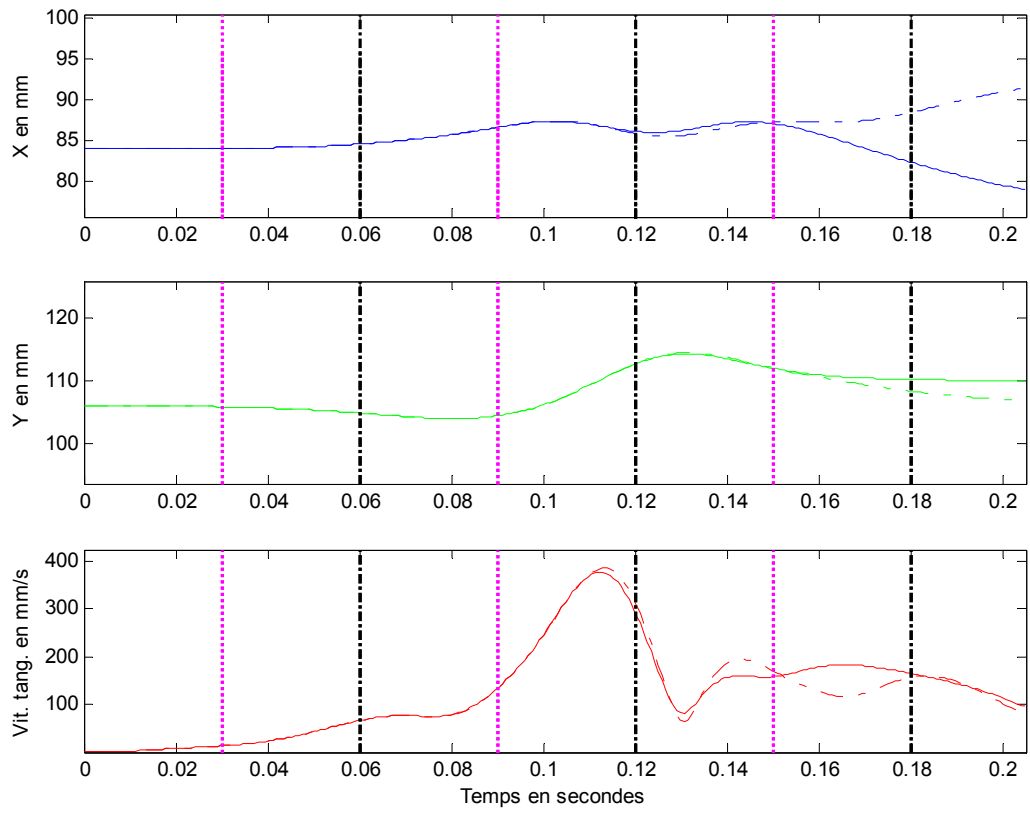

**Figure 7 : Même chose que sur la figure 5, mais pour la condition temporelle 3.**

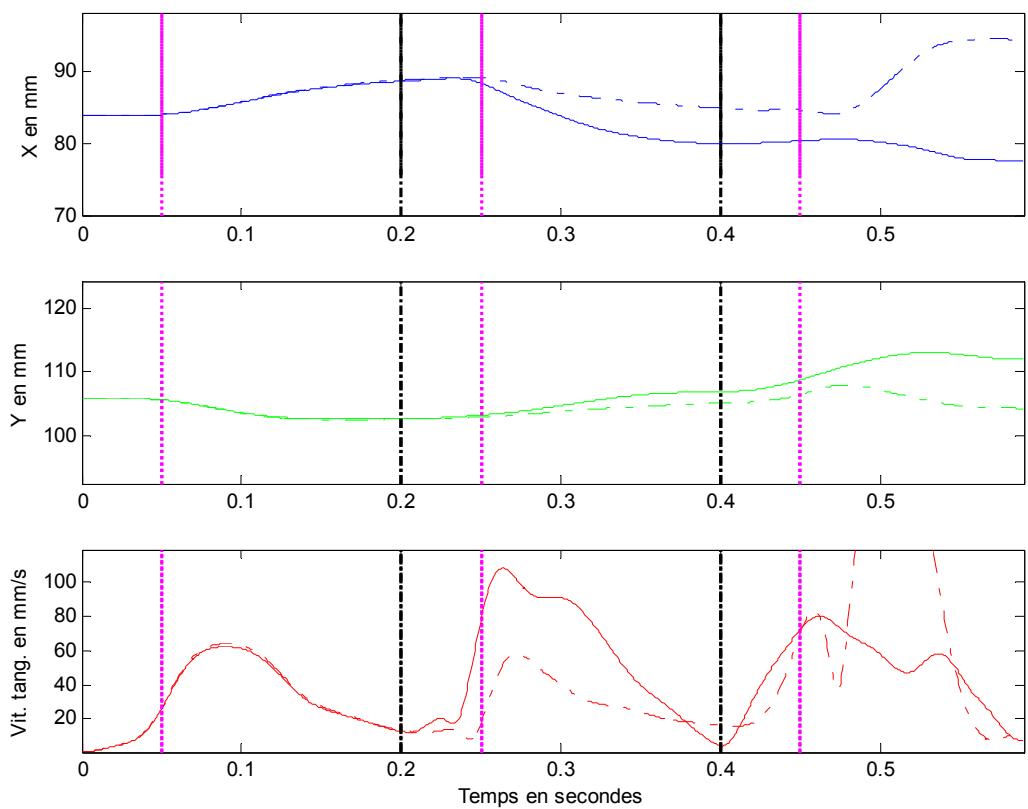

**Figure 8 : Déplacements horizontaux (en haut) et verticaux (au milieu), et vitesse tangentielle (en bas) du point du contour supérieur du modèle représenté à la figure 1 pour les séquences [schwa-a-ɛ-i] (trait continu) et [schwa--a-ɛ-ɔ] (trait tireté) généré par notre modèle de contrôle otpimal. Pour plus de détails voir figure 2**